%% file: main.tex
\documentclass[reprint,a4paper,aps,prl, 
longbibliography,    
floatfix, 
showkeys  
]{revtex4-2}

\input{Preamble/packages}

\begin{document}

\title{Flow topology classification of limit cycles}

\input{Preamble/authors}

\date{\today}

\begin{abstract}
    Recent topological tools offer a powerful way to classify how phases of nonlinear bosonic resonators are organized. Yet, they remain incomplete. In particular, self-sustained oscillations in the form of limit cycles act as robust organizing centers in phase space that are not captured by existing fixed-point-based approaches.
In this work, we extend the flow topology framework for nonlinear resonators to include limit cycles as fundamental topological elements. Using a graph-based construction, we show how periodic attractors impact the global connectivity of phase-space flows. We illustrate the approach with a minimal nonlinear Van der Pol resonator model, where limit cycles coexist with stationary points.
Our results provide a unified topological description of stationary and time-periodic phases in nonlinear bosonic systems, with direct relevance to photonic, superconducting, and optomechanical platforms, and raise new questions on synchronization and the extension of flow topology to the quantum regime.
\end{abstract}

\maketitle

Topological phases of matter reveal physical behavior that is governed by global, quantized invariants rather than local order parameters~\cite{TKNN_Thouless1982,resta2007theory,xiao2010berry,hasan2010colloquium,qi2011topological,bernevig2013topological,ozawa2019topological}. These bulk indices fix quantized response functions that remain robust against disorder and other perturbations, and often manifest through anomalies that reflect violated conservation laws~\cite{frohlich2023gauge}. Classic examples include the quantum Hall effect~\cite{HallEffect_Klitzing1980,TKNN_Thouless1982}, topological insulators protected by time-reversal symmetry~\cite{hasan2010colloquium,qi2011topological}, higher-order phases connected to topology in auxiliary dimensions~\cite{resta2007theory,teo2010topological,kraus2013four,benalcazar2017quantized,petrides2018six,petrides2020higher,kovsata2021second}, and topological effects in quasiperiodic structures~\cite{kraus2012topological,FibonacciPump_Zilberberg2015,zilberberg2021topology}. While these ideas originated in closed and energy-conserving settings, they now extend to open systems where gain and loss play an essential role. Effective ``Non-Hermitian Hamiltonians'' illustrate this trend and highlight the need to generalize notions of topology when spectra become complex and boundary sensitivity increases~\cite{Masahito2020NHphysics,Kunst2021EPs,Bagarello2022NonHermitianMasterEq,FoaTorres2018NHedgestates}. Concepts such as the spectral winding number~\cite{Slager2020SpectralWindingNumber,SpectralWindingNumber,Sato2019SymmetryAndTopology,Fang2020WindingNumberAndCurrent} and non-Bloch band theory~\cite{Yokomizo2019NonBloch,Yokomizo2021BdGNonBloch,Murakami2022NonBlochContinuum,Wang2024NonBlochContinuum,bestler2025non} have been developed to address these challenges and show that topological reasoning remains powerful away from equilibrium.

Moving to many-body bosonic systems, the introduction of nonlinearities forces a departure from the standard program of classifying phases through the spectrum of Hermitian or non-Hermitian quadratic operators. Nonlinear dynamics reshapes the interplay between drive and loss and generates distinct nonequilibrium phases, each with its own linear (quadratic) environment, yet different in their global organization in phase space. Dissipative phase transitions provide a natural arena for this behavior~\cite{Bartolo2016,Fitzpatrick2017,Minganti2018,Soriente2021,Minganti2023,Beaulieu2025}, since nonlinearities and forces conspire to produce multiple NESS, metastability, and critical slowing down across platforms that range from Kerr oscillators~\cite{Soriente2021,eichler2023classical,ChavezCarlos2024,Iyama2024} and Bose Hubbard lattices~\cite{Tomita2017,Vicentini2018} to polariton fluids~\cite{Carusotto2013,Fink2018} and superconducting arrays~\cite{Flurin2017,Wang2019,Burkhart2021,Cai2021,Peyruchat2024,Kollar2024}. To capture the global topological structure of such nonlinear dynamics, we recently introduced  flow topology based on Morse-Smale graphs~\cite{Oshemkov1998,villa2025}. Originally developed to classify watershed networks, this graph index records how attractors, repellors, and saddles connect through the vector field and therefore characterizes the variety and arrangement of dynamical phases in phase space. Transitions between regions with different graph indexes define flow-topology transitions, which correspond to qualitative rearrangements of the NESS landscape, and have been observed in experiments~\cite{villa2025}. This framework also extends to the quantum regime, where its structure persists beyond semiclassical theory~\cite{seibold2025manifestations}.

Our flow topology framework characterizes how stationary solutions organize in the phase space of nonlinear resonators, but it does not account for periodic attractors. Such so-called limit cycles (LCs) are isolated, self-sustained periodic solutions of nonlinear equations, with an oscillation period fixed by the internal balance of forces~\cite{Strogatz2000}. Asymptotic stability and the absence of a preferred time origin make them robust organizing centers of nonlinear dynamics. They arise broadly in out-of-equilibrium platforms, including light-matter ensembles~\cite{Seibold2019, CarlonZambon2020, Marconi2020, Keeling2010, Bhaseen2012b, Lee2013, Piazza2015, Chiacchio2019, Kessler2019,peters2021limit, Kongkhambut2022}, parametric and electromechanical oscillators~\cite{Bello2019a, CalvaneseStrinati2020, Inagaki2021, eichler2023classical}, and superconducting or optomechanical devices~\cite{Metzger2008,Lorch2014a,Bakemeier2015,NavarroUrrios2017,Rudner_2020,PhysRevLett.129.063605}. Because LCs reshape how regions of phase space connect, any topological classification of nonlinear resonator flows remains incomplete without their inclusion. Interestingly, the original graph construction for watershed networks includes elements that represent LCs~\cite{Oshemkov1998}, thus a consistent topological description of resonator flows should be extendable to include them.

\begin{figure*}
    \centering
    \includegraphics[width = 0.8\linewidth]{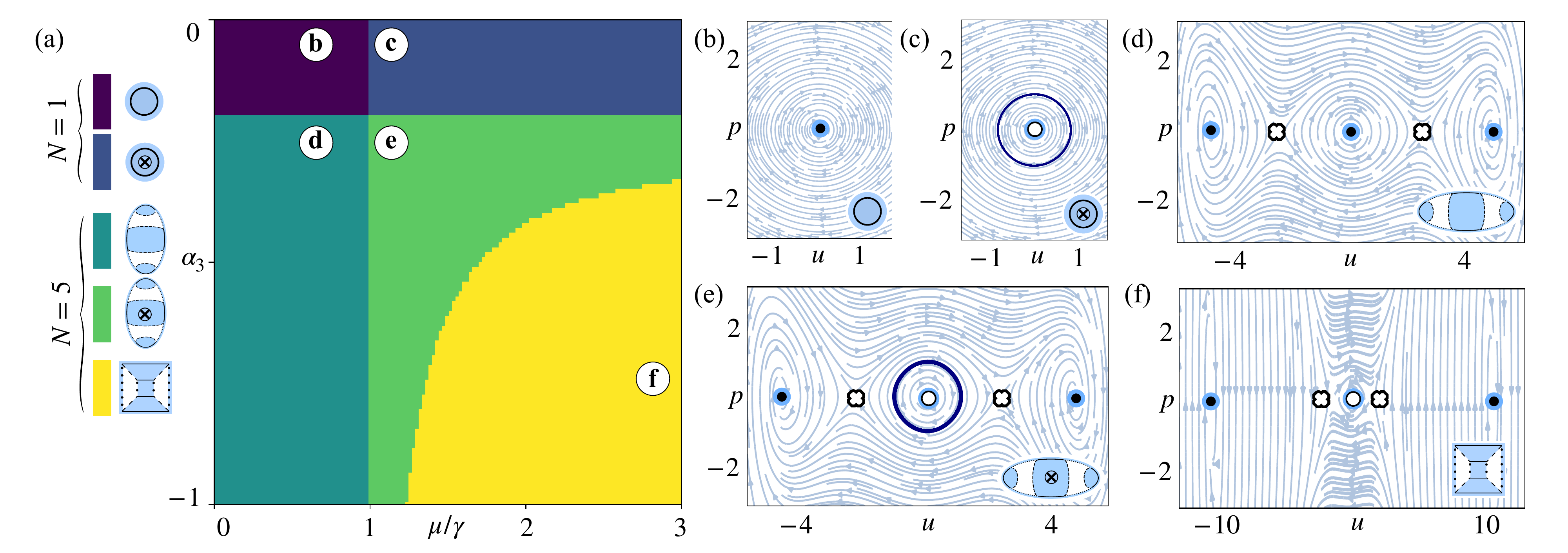}
    \caption{\textit{Vector-flow topology phase diagram}.
(a) Phase diagram of the modified vdP~\eqref{eq:vdp} as a function of $\alpha_3$ and the ration $\mu/\gamma$, with colors indicating distinct vector-flow topologies. Each region corresponds to a different global organization of phase-space trajectories, see legend where $N$ denotes the number of stationary states. 
(b)–(f) Representative phase-space vector fields [cf.~Eqs.~\eqref{eq:vflow}] at the parameter points marked by bullets in panel (a); we observe how fixed points and limit cycles reshape the global flow structure across the diagram. Filled dots, empty circles, and cross markers correspond to attractors, repellers, and saddles, respectively. The color around the dots/circles mark the chirality around the fixed point with blue, red, and white  corresponding to CW, CCW and no winding, respectively. Dark blue lines trace CW-rotating limit cycles. Insets show the associated graph indices characterizing the flow topology in each regime. In all panels, $\omega_0=1$ , $\gamma=0.1$,  $\alpha_5=0.01$.
(b,d) The system does not admit LCs for $\gamma < \mu$. Depending on whether $\alpha_3 \lessgtr -\sqrt{4\alpha_5}$, we either have (b) an elementary flow with only one steady state or (d) a non-elementary flow, with three CW attractors and two saddles. (c,e) As $\gamma > \mu$, the system exhibits a CW attracting LC around the origin (c) as a single feature or (e) coexisting with stationary states. In order for the flow lines of the inner region to converge to the LC, the CW attractor at the origin changes stability and becomes a CW source. (f) As the ratio between $\mu$ and $\gamma$ increases further, the LC is not sustained anymore and vanishes. The source at the origin does not change stability.}
    \label{fig:1}
\end{figure*}

In this work, we extend the flow topology classification of nonlinear bosonic resonators to include limit cycles as genuine topological elements of the phase-space organization. Building on the flow-topology graph framework, we show how periodic attractors modify the global connectivity of flows and give rise to new classes of flow topology beyond fixed-point-based classifications. We demonstrate this extension explicitly by constructing a minimal nonlinear resonator that realizes an extended Van der Pol dynamics, in which stable limit cycles coexist and compete with stationary  states. Our results establish a unified topological description of stationary and time-periodic phases in nonlinear bosonic systems, with direct relevance across photonic, superconducting, and optomechanical platforms. They further raise open questions about how such flow-topological structures persist in the quantum regime and how they constrain synchronization and collective dynamics in many-body bosonic systems.

\begin{figure}
    \centering
    \includegraphics[width = 0.99\linewidth]{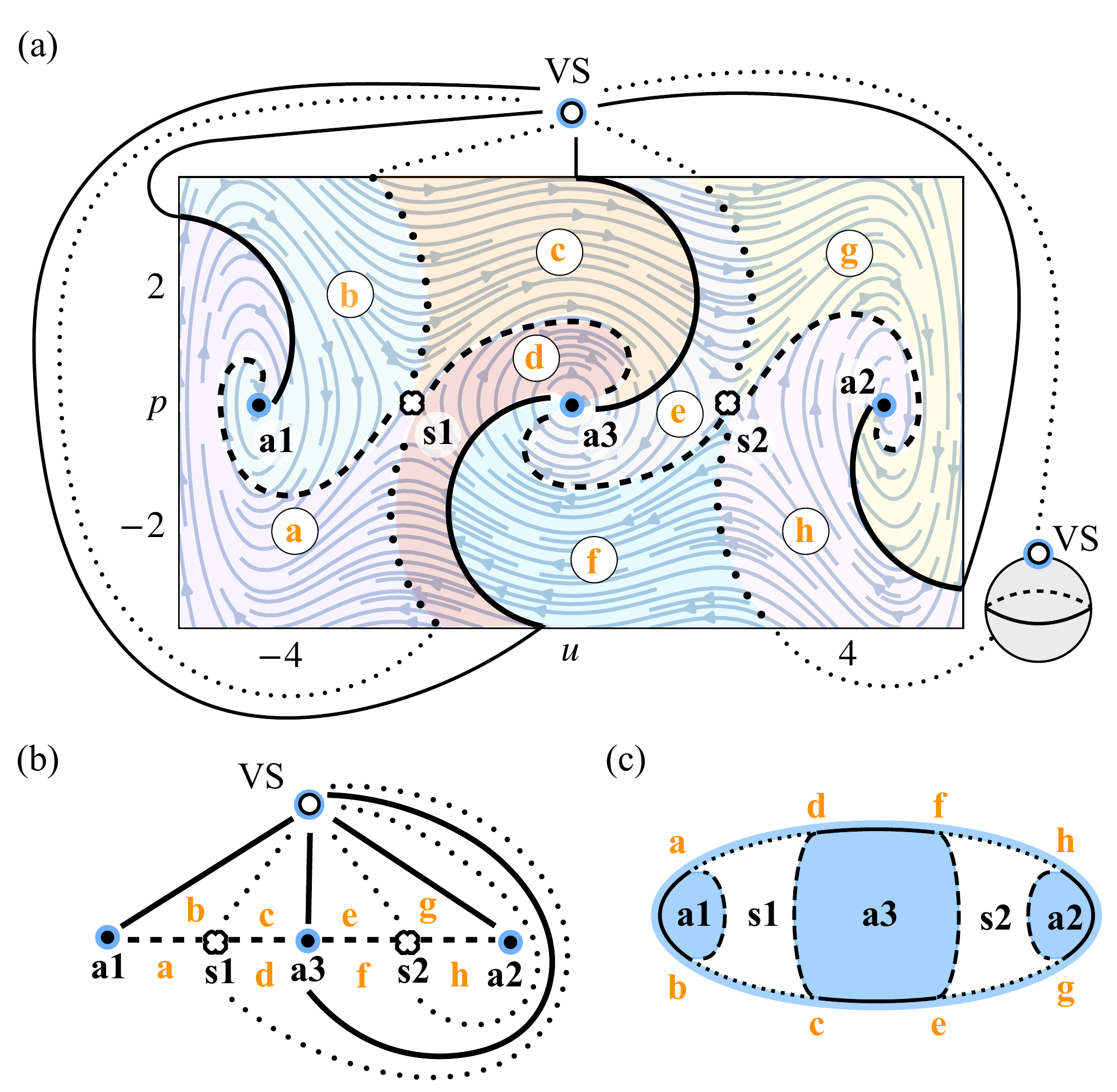}
    \caption{\textit{Topological graph invariant without LCs.} We explain the construction of the vector-flow graph invariant for the flow in Fig.~\ref{fig:1}(d), containing only point-like stationary states, with markers as in Fig.~\ref{fig:1} enumerated by black letters~\cite{Oshemkov1998,villa2025}. (a) Since all flow lines are in-coming from infinity, we can perform a one-point compactification step and map infinity to a virtual source (VS). Thus, the whole vector flow can be viewed as lying on the bottom half of a sphere (inset), with all flow lines descending from the top. The separatrices, special trajectories connecting critical points, partition the surface into regions where all flow lines share the same asymptotic behavior. Starting from the saddle points, we use forward and backward time-evolution to draw the separatrices leading to attractors (dashed lines) and originating from VS (dotted), respectively. In a triangulation step, we add the solid lines as representative trajectories  connecting VS to the attractors (this choice is arbitrary and has no impact on the topology). The surface is now divided in triangular patches, which we highlight with different colors and identify with orange letters. Crucially, this ensemble of separatrices, critical points and faces is topologically robust and takes the name of Morse-Smale Complex (MSC)~\cite{Oshemkov1998}. The separatrices in (a) are simplified for visual purposes~\cite{supmat}. (b) Schematic  representation of the MSC, for better clarity.  (c) The topological planar graph invariant is built as the dual of the MSC. The nodes are the faces, while the lines are the common edges between them, e.g, faces \textbf{a} and \textbf{b} share a solid and a dashed edge, hence nodes \textbf{a} and \textbf{b} are connected by a solid and a dashed line. Each region of the planar graph corresponds to a critical point. The chirality around each critical point is encoded in the coloring: here, we use blue shading to highlight that all attractors are CW. Since the VS is also CW, we draw a blue halo around the graph.}
    \label{fig:2}
\end{figure}

\begin{figure}
    \centering
    \includegraphics[width = 0.99\linewidth]{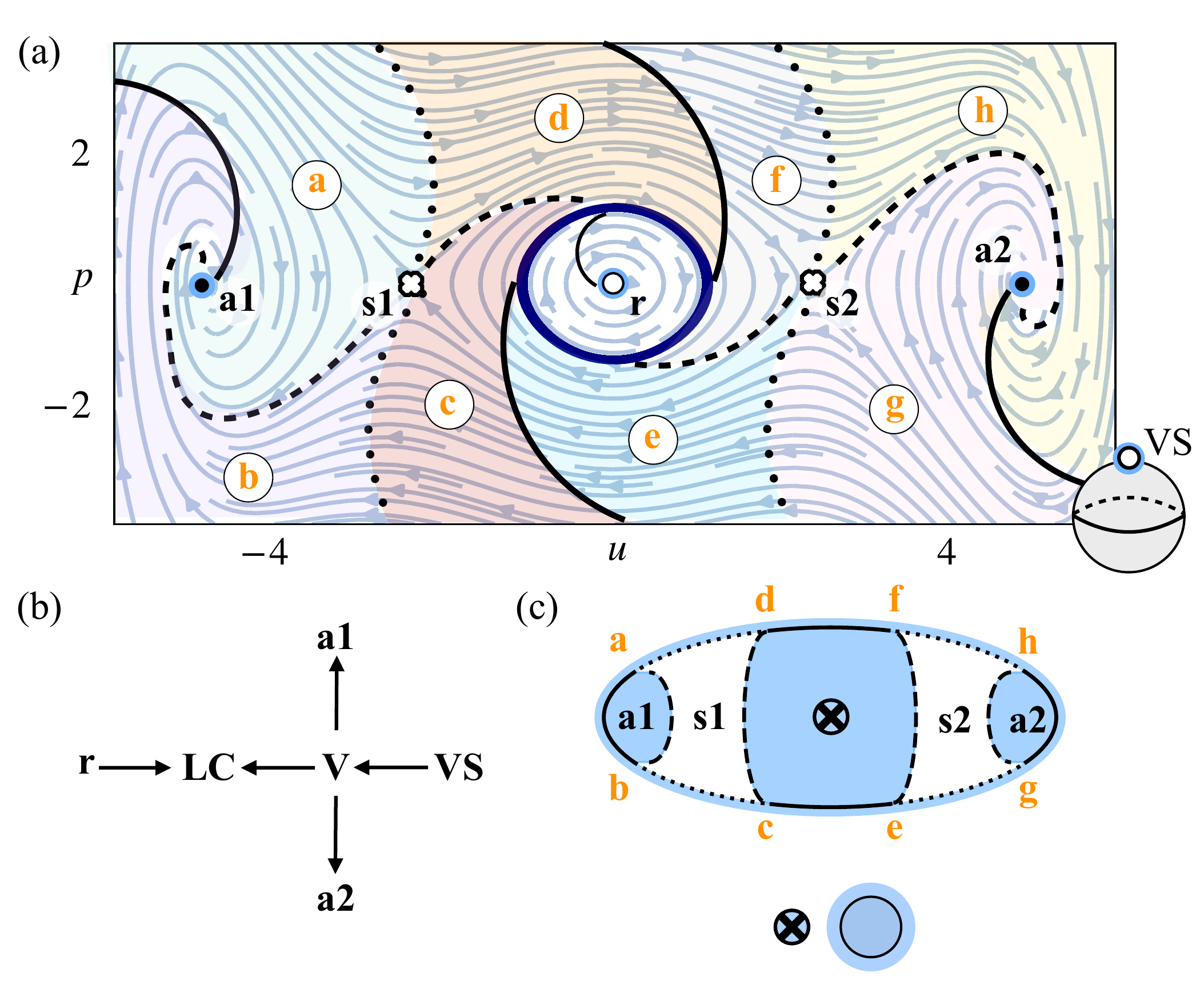}
    \caption{\textit{Extension of the graph invariant formalism to LCs.} We construct the topological graph invariant for the flow in~\autoref{fig:1}(e), where a CW attracting LC is sustained around the origin. (a) We first subdivide the phase space in regions, each delimited by separatrices of three different dashing styles. Here we omit the VS in the drawing. Separatrices are schematized for graphical purposes~\cite{supmat}. (b) Invariant for a flow with a single limit cycle, according to Ref.~\cite{Oshemkov1998}. The LC separates the phase space in an inner ($r$) and an outer region ($V$). The arrows represent flow directionality with respect to the LC: since it is attracting, the inner and outer regions both point to the LC. The fixed-points $a_1$, $a_2$ and $VS$ represent the elementary regions of $V$, with arrows reflecting the flow asymptotic behaviour. Since we only have a single LC, we omit  information on the orientation that the invariant in Ref.~\cite{Oshemkov1998} introduces. (c) Our proposed graph invariant augments the construction from  Ref.~\cite{villa2025} with the additional information of what is inside and outside the LC. We first draw the graph invariant for the outer region, which in this case results in being topologically equivalent to that of ~\autoref{fig:1}(d), when if one replaces the attracting LC with a point-like attractor. We mark the LC region on V with a crossed (double-lined) circle to show that the region inside the LC is repelling (attracting). This uniquely identifies the position of the LC within the graph and the topological behavior of its content. Crucially, the choice of using the colored faces and halo already contain the information on the LC orientation.}
    \label{fig:3}
\end{figure}

%
To demonstrate how to include LCs in the topological classification of nonlinear bosons, we consider a modified van der Pol (vdP) oscillator that includes additional polynomial potential terms in the equation~\cite{del2024limit}
\begin{equation}\label{eq:vdp}
\ddot{x}+\omega_0^2 x+\gamma\dot{x}-\mu(1-x^{2})\dot{x}+\alpha_3 x^{3}+\alpha_5 x^{5}=0\,,
\end{equation}
where $x(t)$ is the oscillator displacement at natural angular frequency $\omega_0, \gamma$ controls the linear damping, $\mu>0$ is the standard vdP parameter that controls the linear gain channel alongside nonlinear damping, and $\alpha_{3,5}$ are cubic and quintic nonlinearities, respectively. 
Equation~\eqref{eq:vdp} can be recast into two coupled first-order differential equations
\begin{align}
\label{eq:vflow}
     \dot{x} &= p\,, \\ \dot{p} &= - \omega_0^2 x +(\mu-\gamma) p -\mu x^2 p  - \alpha_3 x^3   - \alpha_5 x^5 \,, \nonumber
\end{align}
that define a two-dimensional vector flow, where depending on the parameter values, the system can exhibit distinct vector flows, which we, in the following, topologically classify,  see~\autoref{fig:1}.
Stationary states satisfy $\dot{x}=\dot{p}=0$, and  correspond to attractors, saddles, or repellors. For example,  consider the limit where $\mu=0$ and $\gamma,\alpha_{3,5}\neq0$: we have a nonlinear oscillator with quartic and sextic potential terms. The system can, thus, have multiple ($N$) stationary states, see Figs.~\ref{fig:1}(b) and (d). Such flows can be readily topologically-classified with the approach of Ref.~\cite{villa2025} as detailed below, cf.~Fig.~\ref{fig:2}. 
Instead, in the limit of $\mu/\gamma > 1$, we obtain the standard vdP resonator, where the gain channel $-\mu \dot{x}$ drives the oscillator to higher amplitudes, and the nonlinear damping, $\mu x^2\dot{x}$, saturates the motion into a limit cycle, see Figs.~\ref{fig:1}(c) and (e). The LC vanishes as the ratio $\mu/\gamma$ increases and interplays with the nonlinear potentials, see Fig.~\ref{fig:1}(f). The central result of this work is to extend the flow topology framework to systems where LCs coexist with stationary states, cf.~Fig.~\ref{fig:3}. We, thus, obtain a complete phase diagram of the model~\eqref{eq:vdp}, which differs from that presented in Ref.~\cite{del2024limit}, as we base our study solely on vector-flow topology, and not on approximate solution-finding approaches.

The vector flows in Figs.~\ref{fig:1}(b), (d) and (f) can be topologically classified through a graph invariant, which we obtain by extending Morse-Smale vector-flow topology~\cite{Oshemkov1998} to equilibrium and out-of-equilibrium nonlinear resonator dynamics~\cite{villa2025}. We succinctly recap the construction of the graph invariant [see Fig.~\ref{fig:2}]: start from the observation that separatrices partition phase space into regions where all trajectories share the same origin, destination, and orientation. Dashed and dotted separatrices bound elementary regions of the flow, which are further subdivided by solid lines in a triangulation step. The resulting decomposition captures the full connectivity of the flow, including how sources, sinks, and saddles are linked by invariant manifolds, see Fig.~\ref{fig:2}(a) and (b). Taking the dual of this construction yields a planar graph whose faces correspond to flow regions and whose edges encode their adjacency, see Fig.~\ref{fig:2}(c). Two flows are topologically equivalent if and only if their corresponding graphs are isomorphic, meaning that one can be smoothly deformed into the other without rearranging separatrices.

Nonlinear bosonic systems require an extension of this framework to account for rotational flow components~\cite{villa2025}. In such systems, trajectories generically wind clockwise (CW) or counterclockwise (CCW) around stationary states, reflecting the presence of a phase reference set by the drive or, in our nondriven system, by causality~\cite{soriente2020distinguishing,Dumont2024a}. This chirality is a robust property of a physical flow and cannot change under smooth deformations. To incorporate this constraint, each region of the graph is assigned a color that encodes the local flow chirality, with CW and CCW winding represented by distinct colors, while non-rotating regions, corresponding to critically-damped local excitations marked by exceptional points,  remain uncolored~\cite{villa2025}, see Fig.~\ref{fig:2}(c). The chirality associated with an effective source at infinity is encoded separately and encodes the sign of the highest nonlinear potential power. With this addition, the graph invariant provides a complete topological classification of nonlinear resonator vector flows that contain only stationary states. Distinct flow phases correspond to nonisomorphic graphs once both connectivity and region coloring are taken into account, and any change of chirality necessarily signals a topological transition.

The planar graph invariant admits a natural extension to flows that support LCs~\cite{Oshemkov1998}. From a flow topological perspective, an isolated LC plays a role analogous to that of a stationary attractor or source, in that it acts as a terminal object for trajectories in its basin, see Fig.~\ref{fig:3}(a). The key difference is that a LC separates phase space into an inner and an outer region, each supporting its own flow structure. As a result, a LC functions as an interface that links the topological invariants of the two regions it bounds. In this sense, the presence of a LC does not invalidate the graph construction, but enriches it by introducing an additional organizing element that mediates how separatrices connect across regions.

To incorporate this structure into the topological classification~\cite{Oshemkov1998}, the vector flow is first decomposed into elementary regions bounded by invariant curves that are transverse to the flow. Regions that contain a single stationary state are treated as before, while regions containing a single attracting or repelling LC are identified as distinct elementary components. The global flow is then reconstructed by specifying how these regions are connected across their boundaries, with the orientation of trajectories across each boundary fixing the directionality of the graph edges, see Fig.~\ref{fig:3}(b). The orientation of the flow along the LC provides a natural reference that constrains how adjacent regions can be consistently attached. 
Formally, in Ref.~\cite{Oshemkov1998}, the invariant is built with respect to the LC~\autoref{fig:2}(b). The inner and outer regions are identified as $V_{\text{in}}$ and $V_{\text{out}}$. Since here the inner region is elementary (containing a single repelling source), we use only $r$ to identify it and use $V$ for the more complex outer region. Since the LC is attracting, both inner and outer flows converge to it, marked by arrows,  $r \rightarrow \text{LC}$ and $\text{LC} \leftarrow V$. To accommodate the structure of $V$, the directionality of its elementary regions is also highlighted in the invariant: $V \rightarrow a_1$ and $V \rightarrow a_2$ show that $V$ feeds into the attractors, while $V \leftarrow VS$ marks the role of the virtual source. We notice here that, in principle, a more complete representation also requires specifying how the limit cycle winds relative to adjacent saddle regions. Within the standard vector-flow formalism~\cite{Oshemkov1998}, this is indicated by a $\pm$ label on the corresponding graph arrows. On an oriented surface, this reduces to difference in winding relative to the overall chirality of the flow, which in our case of phase-space flow on a disk simplifies to taking winding relative to the chirality of the virtual source. 

As previously discussed, nonlinear bosonic systems need to also account for rotational flow components. Hence, we can enhance the graph from Ref.~\cite{villa2025} to allows for a simple natural extension that  combines the LC chiralities with the stationary state chiralities. Specifically, we augment the colored-graph formalism to retain information on (i) the number of subregions into which the manifold is divided, (ii) their internal topological structure, and (iii) the chirality around each critical point or limit cycle, see~\autoref{fig:3}(c). For a flow containing a single limit cycle, we first construct the graph invariants associated with the inner and outer regions. In this spirit, the upper panel of Fig.~\autoref{fig:3}(c) corresponds to the node $V$ of the invariant in Fig.~\autoref{fig:3}(b). The inner region of the limit cycle can host non-elementary flows, which will globally behave as an attractor/repelling source (depending on the LC's stability). In order to identify where the LC is located in the graph invariant and what is the behaviour of its inner region, we use an additional marker: empty double-lined circle when $V_\text{in}$ is attractive (not relevant in our case), and a crossed circle when $V_\text{in}$ is repelling (relevant in our case). The stability of the limit cycle itself is also visible from the edges of the corresponding graph invariant region, as they are built via separatrices. The chirality is encoded by introducing the same coloring scheme as in Fig.~\autoref{fig:2}. Finally, we complete the picture by showing the topological invariant encoding the structure of the inner region as a legend. This yields a complete topological description of the vector flow that remains invariant under smooth deformations.

Our proposed construction offers significant advantages: (i) it immediately provides information both on the global and on the local structure of the flow instead of hiding the latter (especially for cases where a LC contains complex structures such as saddle regions); (ii) it uniquely identifies the position of each invariant element and the directionality of the flow in its surroundings; (iii) it has an in-built specification of the orientation of the LC. The last point becomes particularly useful when additional LCs are involved. Instead, in the graphical representation of Ref.~\cite{Oshemkov1998}, attractors, sources and LCs are visually not uniquely specified (all limit cycles are called $S$-atoms, while all elementary regions are $A$-atoms). In~\autoref{fig:3}(b), we individually identify each critical point/LC to make the comparison with our proposed graph clearer\cite{supmat}. This change in notation preserves the topological content of the invariant and better reflects our requirement to uniquely associate each critical point with its position in the graph, a feature that is essential in nonlinear bosonic systems.

We extend flow-topological classification of driven-dissipative nonlinear resonators to include limit cycles as fundamental topological objects. By generalizing the Morse–Smale graph construction beyond stationary fixed points, we show how self-sustained oscillations reorganize phase-space connectivity and generate flow-topological phases that cannot be captured by fixed-point-based indices alone. Using a minimal extension of the Van der Pol resonator, we explicitly demonstrate how the emergence and bifurcation of limit cycles modify global flow topology and induce flow-topology transitions associated with qualitative rearrangements of dynamical basins. This extension opens several natural directions. Classically, it provides a systematic framework to analyze synchronization, mode locking, and multicycle dynamics in coupled resonator networks. Quantum mechanically, it raises fundamental questions about how periodic attractors appear in quantum steady states and trajectories, and whether their flow-topological signatures persist under strong fluctuations. More broadly, incorporating limit cycles advances flow topology toward a fully dynamical classification of nonequilibrium phases, connecting topology, nonlinear dynamics, and open quantum systems.

While submitting this manuscript, we became aware of a similar work~\cite{Gomez2025}.

\section*{Acknowledgements}
We acknowledge funding from the Deutsche Forschungsgemeinschaft (DFG) through project numbers 449653034, 521530974, and 545605411, as well as via SFB 1432 (project number 425217212). We also acknowledge support from the Swiss National Science Foundation (SNSF) through the Sinergia Grant No. CRSII5\_206008/1.\\

\bibliographystyle{apsrev4-2}

\setcitestyle{super,compress}

\bibliography{bib.bib}

\end{document}

%% file: Preamble/packages.tex
\usepackage{mathtools}       
\usepackage{amssymb}         
\usepackage{dsfont}          
\usepackage{bm}              
\usepackage{physics}         

\usepackage{siunitx}         

\usepackage[T1]{fontenc}     
\usepackage[utf8]{inputenc}  
\usepackage{lmodern}         

\usepackage{xcolor}          

\usepackage{graphicx}        
\usepackage{float}           

\usepackage{microtype}       

\usepackage[colorlinks = true, linkcolor = blue, citecolor = blue, urlcolor = blue]{hyperref}

%% file: Preamble/authors.tex

\newcommand{\UKON}{Department of Physics, University of Konstanz, 78457 Konstanz, Germany}

\author{Thomas Mutschler}
\affiliation{\UKON}

\author{Greta Villa}
\affiliation{\UKON}

\author{Oded Zilberberg}
\affiliation{\UKON}